# Infrared hyperbolic metasurface based on nanostructured van der Waals materials

*Peining Li*[1], *Irene Dolado*[1], *Francisco Javier Alfaro-Mozaz*[1], *Felix Casanova*[1,2], *Luis E. Hueso*[1,2], *Song Liu*[4], *James H. Edgar*[4], *Alexey Y. Nikitin*[1,2,3], *Saül Vélez*[1,†], *Rainer Hillenbrand*[2,5*]

1 CIC nanoGUNE, 20018, Donostia-San Sebastián, Spain.
2 IKERBASQUE, Basque Foundation for Science, 48013 Bilbao, Spain.
3 Donostia International Physics Center (DIPC), 20018 Donostia-San Sebastián, Spain
4 Dept. of Chemical Engineering, Kansas State University, Manhattan, 66506 Kansas, USA.
5 CIC NanoGUNE and UPV/ EHU, 20018, Donostia-San Sebastián, Spain
† Present Address: Department of Materials, ETH Zürich, 8093 Zürich, Switzerland

*Correspondence to: r.hillenbrand@nanogune.eu

**Metasurfaces with strongly anisotropic optical properties can support deep subwavelength-scale confined electromagnetic waves (polaritons) that promise opportunities for controlling light in photonic and optoelectronic applications. We develop a mid-infrared hyperbolic metasurface by nanostructuring a thin layer of hexagonal boron nitride supporting deep subwavelength-scale phonon polaritons that propagate with in-plane hyperbolic dispersion. By applying an infrared nanoimaging technique, we visualize the concave (anomalous) wavefronts of a diverging polariton beam, which represent a landmark feature of hyperbolic polaritons. The results illustrate how near-field microscopy can be applied to reveal the exotic wavefronts of polaritons in anisotropic materials, and demonstrate that nanostructured van der Waals materials can form a highly variable and compact platform for hyperbolic infrared metasurface devices and circuits.**

Optical metasurfaces are thin layers with engineered optical properties (described by the effective permittivities in the two lateral directions), which are obtained by lateral structuring of the layers (*1-3*). Applications include flat lenses, high-efficiency holograms, generation of optical vortex beams and manipulation of polarization state of light (*1-5*). With metallic metasurfaces one can also control the properties of surface plasmon polaritons (SPPs, electromagnetic waves arising from the coupling of light with charge oscillations in the metasurface) propagating along the metasurface. The near-field enhancement and confinement provided by SPPs is another effective means for controlling the phase and polarization of transmitted light, or the thermal radiation emitted from the metasurface (*2, 3*). Metasurfaces can also be used to control the properties of SPPs in nanophotonic circuits and devices, for applications such as unidirectional excitation of SPPs, modulation of SPPs or two-dimensional (2D) spin optics (*2, 6, 7*).

Recently, hyperbolic metasurfaces (HMSs) were predicted — uniaxial metasurfaces where the two effective in-plane permittivities $\varepsilon_{\text{eff},x}$ and $\varepsilon_{\text{eff},y}$ have opposite signs (*2, 6*). In such materials the SPPs exhibit a hyperbolic in-plane dispersion, i.e. the isofrequency surface in wavevector space describes open hyperboloids (*2, 6, 8-13*). Consequently, the polaritons on HMSs possess an extremely anisotropic in-plane propagation (i.e. different wavevectors in different lateral directions). This behavior leads to remarkable photonic phenomena. For example, the wavefronts of a diverging polariton beam emitted by a point-like source can exhibit a concave curvature (*6, 8*), in stark contrast to the convex wavefronts in isotropic materials. Further, the large wavevectors (limited only by the inverse of the structure size) yield a diverging, anomalously large photonic density of states, which can be appreciably larger than that of isotropic SPPs (*2, 8*). Such polariton properties promise intriguing applications including planar hyperlenses (*2, 8*), diffraction-free polariton propagation (*6, 13*), engineering of polariton wavefronts (*6*), 2D topological transitions (*8*) and super-Coulombic optical interactions (*10*).

HMSs could be created artificially by lateral structuring of thin layers of an isotropic material (*2, 6, 13*). Alternatively, 2D materials with natural in-plane anisotropy, i.e. black phosphorous, could represent a natural class of HMSs (*14-16*). However, only few experimental studies at microwave (*11, 12*) and visible frequencies (*13*) have been reported so far, demonstrating only weakly confined SPPs on structured metal surfaces. Artificial HMSs at mid-infrared and terahertz frequencies (corresponding to energies of molecular vibrations, and thermal emission and absorption) have not been realized yet, while visualization of the diverging concave wavefronts of deeply subwavelength-scale confined hyperbolic polaritons on either artificial or natural HMSs has been elusive.

Here we propose, design and fabricate a mid-infrared HMS by lateral structuring of thin layers of the polar van der Waals (vdW) material hexagonal boron nitride (h-BN). In contrast to metal layers, they support strongly volume-confined phonon polaritons (quasi-particles formed by coupling of light with lattice vibrations) with remarkably low losses, thus representing a suitable basis for mid-infrared HMSs.

h-BN is a polar vdW (layered) crystal, thus possessing a uniaxial permittivity (*17-23*). It has a mid-infrared Reststrahlen band from 1395 – 1630 cm$^{-1}$ (*24*), where the in-plane permittivity is negative and isotropic, $\varepsilon_{hBN,x} = \varepsilon_{hBN,y} = \varepsilon_{hBN,\perp} < 0$, while the out-of-plane permittivity is positive, $\varepsilon_{hBN,z} = \varepsilon_{hBN,\parallel} > 0$. As a result, the phonon polaritons in natural h-BN exhibit an out-of-plane hyperbolic dispersion, while in-plane dispersion is isotropic (*17-23*). The isotropic (radial) propagation of the conventional hyperbolic phonon polaritons (HPhPs) in a natural h-BN layer is illustrated with numerical simulations (Fig 1A-D) where a dipole above the h-BN layer launches essentially the fundamental slab mode M0 (*17-19*). To turn the h-BN layer into an in-plane HMS, a grating structure is patterned consisting of h-BN ribbons of width *w* that are separated by air gaps of width *g* (Fig. 1E). Due to the anisotropic permittivity of h-BN, however, the h-BN grating represents a biaxial layer exhibiting three different effective permittivities ($\varepsilon_{\text{eff},x} \neq \varepsilon_{\text{eff},y} \neq \varepsilon_{\text{eff},z}$), in contrast to uniaxial metal gratings where $\varepsilon_{\text{eff},x} = \varepsilon_{\text{eff},y} \neq \varepsilon_{\text{eff},z}$ (which can be considered as canonical HMS structures at visible frequencies). These designed biaxial layers can support highly confined polaritons with in-plane hyperbolic dispersion. They could be considered as Dyakonov-like polaritons (*25*), which are similar to Dyakonov waves on biaxial dielectric materials (*26*).

We first applied effective medium theory (*25*) to determine the parameters required for the grating to function as an HMS. The effective permittivities were calculated as a function of ribbon and gap widths, *w* and *g* (figs. S1 and S2), according to

$$\varepsilon_{eff,x} = 1 / \left[ \frac{1-\xi}{\varepsilon_{hBN,\perp}} + \xi \right] \quad (1)$$

$$\varepsilon_{eff,y} = \varepsilon_{hBN,\perp}(1 - \xi) + \xi \quad (2)$$

$$\varepsilon_{eff,z} = \varepsilon_{hBN,\parallel}(1 - \xi) + \xi \quad (3)$$

where $\xi = g / (w + g)$ is the so called filling factor. We find that the condition for well-pronounced in-plane hyperbolic dispersion, $-10 < \text{Re}(\varepsilon_{\text{eff},y})/\text{Re}(\varepsilon_{\text{eff},x}) < 0$, can be fulfilled in the frequency range 1400 – 1500 cm$^{-1}$ for filling factors in the order of $\xi = 0.5$. For example, at a frequency $\omega = 1425$ cm$^{-1}$ ($\varepsilon_{\text{h-BN},\perp} = -22.2 + 0.9i$ and $\varepsilon_{\text{h-BN},\parallel} = 2.6$) we obtain $\varepsilon_{\text{eff},x} = 3.7$, $\varepsilon_{\text{eff},y} = -15.2 + 0.6i$ and $\varepsilon_{\text{eff},z} = 2.1$ for $w = 70$ nm and $g = 30$ nm, which corresponds to a grating structure that can be fabricated by electron beam lithography and etching. Note that the same effective in-plane permittivities for a metal grating ($\text{Re}(\varepsilon_{\text{metal}}) < -1000$ at mid-infrared frequencies) would require 100 nm wide ribbons separated by less than 1 nm wide gaps (fig.S1), thus strongly challenging nanofabrication technologies. Alternatively, one could use doped semiconductors and (isotropic) polar crystals, where the permittivities in the mid-infrared and terahertz spectral range are similar to that of h-BN (*27, 28*). However, the losses in doped semiconductors are typically larger than in h-BN. Phonon polaritons in polar crystals such as SiC exhibit low losses similar to h-BN, but thin layers are difficult to grow without defects, the later increasing losses (*27, 28*). h-BN thus represents a promising material for the experimental realization of grating-based HMSs, owing to the easy fabrication of high-quality single-crystalline thin layers.

To verify the HMS design parameters, we carried out numerical simulations (Note S3) of dipole-launched polaritons on a 20 nm thick h-BN grating, using $w = 70$ nm, $g = 30$ nm and bulk h-BN permittivities at $\omega = 1425$ cm$^{-1}$ (Figs, 1F, G). The typical hyperbolic polariton rays can be seen in the intensity image (Fig. 1F), while the real part of the electric field distribution reveals the concave polariton wavefronts (Fig. 1G). The formation of these wavefronts arises from an interference phenomenon of polaritons that propagate with a direction-dependent wavelength (determined by the hyperbolic isofrequency curves) at a given frequency. Fourier transform (FT) of Fig. 1G corroborates the polaritons´ in-plane hyperbolic dispersion as well as large wavevectors (deep subwavelength-scale confinement), and thus that the biaxial grating structure functions as a HMS. We repeated the numerical simulation of dipole-launched polaritons on a 20 nm thick biaxial layer with the corresponding effective permittivities (fig. S3). Excellent quantitative agreement is found with the simulated near-field distribution above the grating structure.

The numerical simulations also show that the wavevector (and thus the confinement) of the hyperbolic metasurface phonon polaritons (HMS-PhPs) is increased compared to that of the conventional HPhPs on the natural h-BN layer (Fig. 1D). They further demonstrate that the HMS-PhP propagation depends on frequency (Figs. 1J, K and L), and that it can be tuned by varying the structure size (Figs. 1F, I and J).

For an experimental demonstration of our proposed metasurface, we etched a 5 μm × 5 μm size grating (schematics in Fig. 2A, topography image in Fig. 2B) into a 20-nm-thick exfoliated flake of isotopically enriched (*29-31*) low-loss h-BN (Note S1). The ribbon and slit widths are 75 and 25 nm, respectively (fig. S4). Near-field polariton interferometry was used to show that the gratings support polaritons (*17, 18*). The metallic tip of a scattering-type scanning near-field microscope (s-SNOM), acting as an infrared antenna, concentrates an illuminating infrared beam to a nanoscale spot at the tip apex, which serves as a point source to launch polaritons on the sample (Fig. 2A). The tip-launched polaritons reflect at sample discontinuities (such as edges and defects), propagate back to the tip and interfere with the local field below the tip. Recording the tip-scattered field (amplitude signals $s$ in our case, Note 2) as a function of the tip position yields images that exhibit interference fringes of $\lambda_p/2$ spacing, where $\lambda_p$ is the polariton wavelength (*17, 18*).

Figure 2C shows the polariton interferometry images of our sample measured at four different frequencies. On both the grating (HMS) and the surrounding (unpatterned) h-BN flake polariton fringes are observed. On the grating we see fringes only parallel to the horizontal HMS boundary, which could be explained by a close-to-zero reflection at the left and right grating boundaries, or more interestingly, by the absence of polariton propagation in *x*-direction (horizontal), the latter being consistent with hyperbolic polariton dispersion. We further observe a nearly twofold reduced fringe spacing on the grating, $d_{HMS}$, compared to that of the unpatterned flake, $d_{h\text{-}BN}$ (see line profiles in Fig. 2D), indicating superior polaritonic field confinement on the grating. For further analysis (see also fig. S5), we compare in Figs. 2F and G the experimental polariton wavevectors $k = 2\pi/\lambda_p = \pi/d$ (symbols) with isofrequency curves of the calculated polariton wavevectors (solid lines). The calculation predicts a hyperbolic isofrequency curve for the grating polaritons, where the wavevector in *y*-direction is increased by nearly a factor of two, which quantitatively matches well our experimental observation. Although the polariton confinement increases on the HMS, the calculated relative propagation length (often used as Figure of Merit (*17, 31*), FOM = $k/\gamma$ with $k$ and $\gamma$ being the real and imaginary part of the complex wavevector $K$) and polariton lifetime remain nearly the same (fig. S6). Experimentally, however, the FOM and lifetime on the metasurface are reduced by more than 35 % as compared to that of polaritons on the unpatterned flake, which we attribute to polariton losses caused by fabrication imperfections leading, for example, to polariton scattering (figs. S7 and S8).

The near-field images provide experimental support that the h-BN grating´s function as an in-plane HMS. However, they do not reveal anomalous (concave) wavefronts such as the ones observed in Fig. 1. This can be understood by considering that only the polaritons propagating perpendicular to the metasurface boundary are back-reflected to the tip and thus recorded (Fig. 2E). The anomalous wavefronts are the result of interference of hyperbolic polaritons propagating in all allowed directions. To obtain real-space images of the anomalous wavefronts, we performed near-field imaging of HMS polaritons emerging from a nanoscale confined source located directly on the sample.

For nanoimaging of the polariton wavefronts we used antenna-based polariton launching (*32*). A gold rod acting as an infrared antenna was fabricated on top of the sample studied in Fig. 2 (illustration in Fig. 3A, topography image in Fig. 3B). Illumination with *p*-polarized IR light excites the antenna, yielding nanoscale concentrated fields at the rod extremities, which launch polaritons on either the metasurface or the unpatterned flake (upper and bottom parts of Figs. 3B and C, respectively). The polariton field propagating away from the antenna, $E_p$, interferes with the illuminating field $E_{in}$, yielding interference fringes on the sample (illustrated by solid lines in Fig. 3B) (*19*). This pattern is mapped by recording the field scattered by the metal tip of the s-SNOM, while the sample is scanned. Since $E_{in}$ is constant (i.e. independent of sample position), the interference pattern observed in the amplitude image directly uncovers the spatial distribution of the polariton field $E_p$ and thus the polariton wavefronts (note that retardation in first order approximation can be neglected owing to the much shorter polariton wavelength compared to the illuminating photon wavelength (*19*)).

Figure 3C shows a near-field image of the antenna on the h-BN sample at $\omega = 1430$ cm$^{-1}$. In the lower part of the image we clearly observe the conventional (convex) polariton fringes, which are caused by antenna-launched HPhPs on the unpatterned h-BN layer (see also fig. S9). The upper part of the image, in striking contrast, exhibits anomalous polariton fringes emerging from the rod´s upper extremity. They clearly reveal the concave wavefronts of a diverging polariton beam on the h-BN grating. The image thus provides clear experimental visualization of in-plane hyperbolic polaritons, and thus verifies the grating´s function as a HMS. By varying the illumination frequency, the anomalous polariton wavefronts can be tuned (Fig. 4A) from smoothly concave (diffraction-free polariton propagation (*6*)) at $\omega = 1435$ cm$^{-1}$ to a wedge-like shape at $\omega = 1415$ cm$^{-1}$. We corroborate the experimental near-field images with numerical simulations shown in Fig. 4B, which indeed show an excellent agreement, particularly regarding the fringe spacing and curvatures.

For simplicity we did not include the metallic tip into the calculations, which in the experiment launches polaritons simultaneously with the gold antenna. Analogously to Fig. 2, the tip-launched polaritons reflect at the boundary between the h-BN grating and the unpatterned h-BN flake, yielding horizontal fringes in the experimental near-field images (marked by arrows in Fig. 4A) that are not seen in the simulated images (Fig. 4B). Interestingly, the tip-launched polaritons are weakly reflected at the gold antenna, thus not producing disturbing interference with the antenna-launched polaritons. The weak polariton reflection at metal structures on top of h-BN samples is consistent with former observations and could be explained by the hyperbolic polaritons propagating through the h-BN underneath the metal structure (*19*). For further details on tip-launched vs. antenna-launched polaritons see also fig. S10.

To determine the in-plane HMS-PhP wavevectors, we performed a FT of the experimental near-field images of Fig. 4A (fig. S11). The FTs are shown in Fig. 4C (see also fig. S3), clearly revealing hyperbolic features. They exhibit an excellent agreement with the numerically calculated hyperbolic dispersions of HMS-PhPs (white dashed lines, see Note S3), thus further corroborating that the real-space images in Fig. 4A reveal the wavefronts of in-plane hyperbolic polaritons. Figure 4c also verifies the large polariton wavevectors *k* achieved with our HMS (for example, $k = (k_x^2 + k_y^2)^{0.5}$ > $20k_0$ at *ω* = 1425 cm$^{-1}$, with $k_0$ being the photon wavevector), which are significantly larger than that of the SPPs on metal-based HMSs ($k < 3k_0$) (*12, 13*). In principle, the wavevectors predicted theoretically (Fig. 1H) and observed experimentally (Fig. 4C) could be larger, amounting up to $70k_0$, which is their theoretical limit given by the grating period. We explain the non-visibility of such large wavevectors by wavevector distribution of near fields of the 250-nm-wide antenna, which exhibits dominantly small wavevectors below $28k_0$. For that reason, large-wavevector polaritons on the HMS are only weakly excited and masked by small-wavevector polaritons. To increase the excitation and relative weight of large-wavevector HMS-PhPs, we suggest to fabricate launching structures of smaller dimensions. In numerical calculations, indeed, we can observe larger polariton wavevectors (and thus more pronounced polariton rays) by simply placing the polariton-launching dipole closer to the HMS surface (see fig. S12). Note that we can also see a second (much weaker) hyperbolic feature at larger *k* values. We assign them to tip-launched polaritons that weakly reflect at the gold rod (fig. S13). Because of their weak reflection, they are barely recognized in the real-space images of Fig. 4A. More interestingly, the FTs of the near-field images verify that the opening angle *θ* of the hyperboloids decreases with increasing frequency (fig. S14). This has been predicted in numerous previous theoretical and numerical studies (*6, 8*), and now can be directly observed in experimental data.

Considering that fabrication of nanoscale gratings by electron beam lithography and etching is a widely applicable technique, we envision HMSs based on other vdW materials ($MoS_2$, $Bi_2Se_3$, etc.) or multilayer graphene samples, as well as on thin layers of polar crystals (SiC, quartz, etc.) or low-loss doped semiconductors (*27, 28*). Combinations of different materials could lead to HMSs covering the entire range from mid-IR to THz frequencies. The combined advantage of strong polariton-field confinement, anisotropic polariton propagation, tunability by geometry and gating, as well as the possibility of developing vdW heterostructures (*33*), could open new exciting possibilities for flatland infrared/thermal and optoelectronic applications, such as infrared chemical sensing, planar hyperlensing (*8, 18, 21*), exotic optical coupling (*10*), or manipulation of near-field heat transfer (*2, 8*). Real-space wavefront nanoimaging of in-plane hyperbolic polaritons, as demonstrated in our work, will be an indispensable tool for verifying novel design principles and for quality control.

**Acknowledgements**

The authors acknowledge support from the European Commission under the Graphene Flagship (GrapheneCore1, Grant no. 696656), the Marie Sklodowska-Curie individual fellowship (SGPCM-705960), the Spanish Ministry of Economy and Competitiveness (national projects MAT2015-65525-R, MAT2015-65159-R, FIS2014-60195-JIN, MAT2014-53432-C5-4-R, FIS2016-80174-P, MAT2017-88358-C3-3-R), the Basque government (PhD fellowship PRE-2016-1-0150), the Regional Council of Gipuzkoa (Project No. 100/16), and the Department of Industry of the Basque Government (ELKARTEK project MICRO4FA). Further, support from the Materials Engineering and Processing program of the National Science Foundation, award number CMMI 1538127, and the II−VI Foundation is greatly appreciated. R. Hillenbrand is co-founder of Neaspec GmbH, a company producing scattering-type scanning near-field optical microscope systems, such as the one used in this study. The remaining authors declare no competing financial interests. Data are available upon request to P.L and R.H.

Supplementary Materials

Materials and Methods (Note S1 to S3)

Figs. S1 to S14

References (34-36)

**Figure caption**

**Fig. 1. Dipole-launching of h-BN phonon polaritons.** (**A**) Schematics of dipole launching of phonon polaritons on a 20-nm-thick h-BN flake. (**B**) Simulated magnitude of the near-field distribution, $|E|$. (**C**) Simulated real part of the near-field distribution, $E_z$. (**D**) Absolute value of the Fourier transform (FT) of panel (**C**). $k_x$ and $k_y$ are normalized to photon wavevector $k_0$. (**E**) Sketch of dipole launching of phonon polaritons on a 20-nm-thick h-BN HMS (ribbon width $w = 70$ nm, gap width $g = 30$ nm). **F**) Simulated magnitude of the near-field distribution, $|E|$. (**G**) Simulated real part of the near-field distribution, $E_z$. (**H**) Absolute value of the FT of panel (**G**). The revealed FT features of dipole-launched polaritons can be well fitted by a hyperbolic curve (white dashed lines). (**I** to **L**) Simulated magnitude of near-field distribution for HMSs with different gap sizes and operation frequencies. The grating period $w+g$ is fixed to 100 nm in all simulations. The white arrows in (**B**) and (**F**) display the simulated power flow.

**Fig. 2. Polariton-interferometry imaging of phonon polaritons on a 20-nm-thick h-BN HMS.** (**A**) Illustration of the near-field polariton interferometry experiment. (**B**) Topography image of the h-BN HMS (nominal grating parameters are $w \approx 75$ nm and $g \approx 25$ nm, see also fig. S4). (**C**) Near-field images (amplitude signal $s$) recorded at four different frequencies. (**D**) Signal profiles along the solid (vertical) and dashed (horizontal) white lines in (**C**), respectively. (**E**) Illustration of the polariton interferometry contrast mechanism. The tip launches phonon polaritons on the HMS (indicated by simulated near fields). The vertically propagating polariton reflect at the lower horizontal boundary (indicated by black arrow) and interfere with the local field underneath the tip. Tip-launched polaritons propagating in other directions (e.g. in direction of the red arrows) cannot be probed by the tip. (**F** and **G**) Experimental (symbols) and numerically calculated (solid lines) wavevectors of phonon polaritons on the unpatterned flake and the HMS, respectively. The line colors indicate the frequency $\omega$ according to (**D**).

**Fig. 3. Wavefront imaging of antenna-launched HMS-PhPs**. (**A**) Illustration of the experiment. (**B**) Topography image. The lines illustrate wavefronts of HMS-PhPs on the HMS or phonon polaritons on the unpatterned flake, respectively. (**C**) Near-field image measured at $\omega$ = 1430 cm$^{-1}$, clearly revealing concave wavefronts of HMS-PhPs emerging from the rod´s upper extremity.

**Fig. 4. Frequency dependence of HMS-PhP wavefronts**. (**A** and **B**) Experimental and calculated near-field distribution of HMS-PhPs launched by the antenna at three different frequencies. Black arrows in (**A**) indicate the fringes of polaritons launched by the tip and reflected at the boundary between the HMS and the unpatterned flake. (**C**) Absolute value of FT of the images of (**A**). White dashed lines represent the numerically calculated isofrequency curves of antenna-launched HMS-PhPs. The features in the gap between the hyperbolic isofrequency curves correspond to the FT of the antenna fields that are not coupled to polaritons (*32*), analogous to the central circular feature in the FT of the dipole-launched HMS-PhPs (Fig. 1H). The light-blue circle at $\omega$ = 1425 cm$^{-1}$ marks a pixel whose value is still above the noise floor. The largest polariton wavevectors thus amount to about $k = (k_x^2 + k_y^2)^{0.5} = ((15k_0)^2 + (15k_0)^2)^{0.5} > 20k_0$.

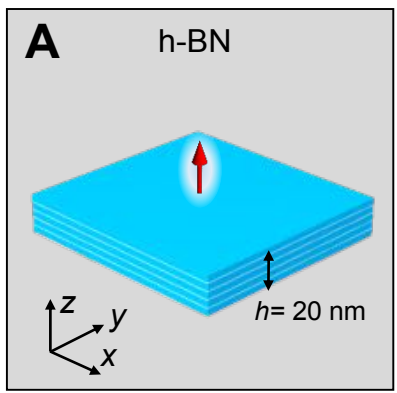 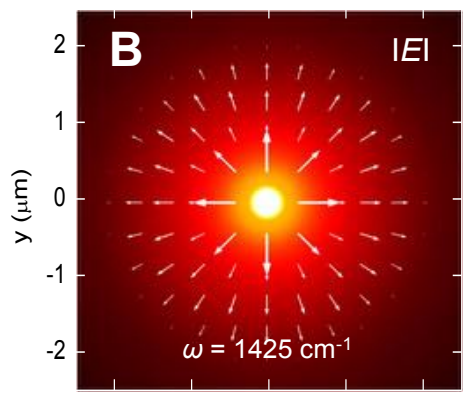 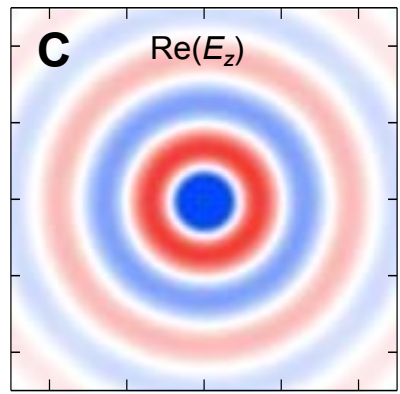 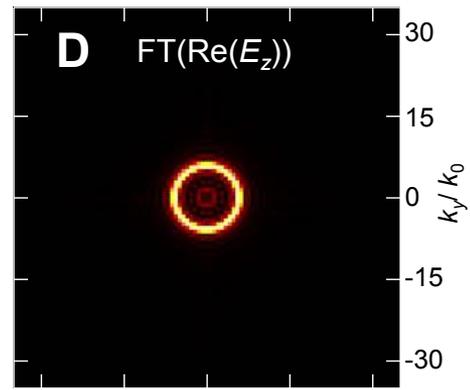
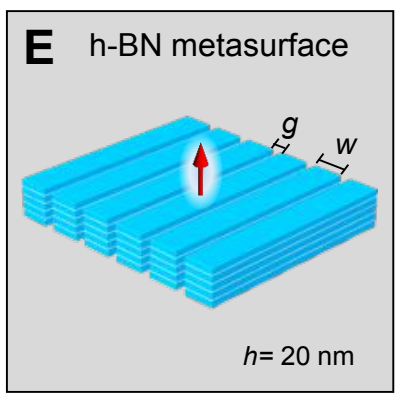 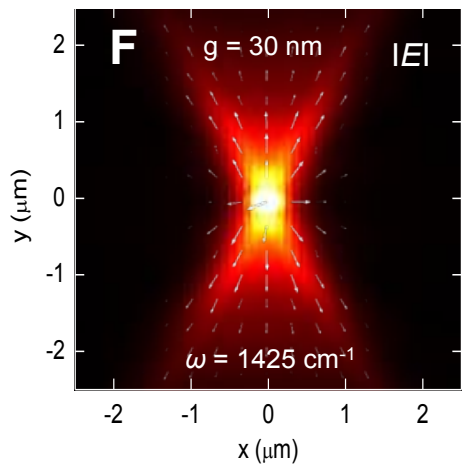 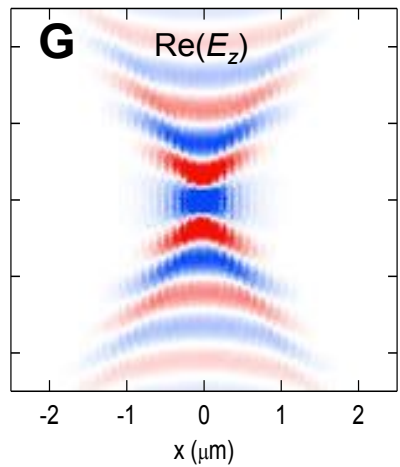 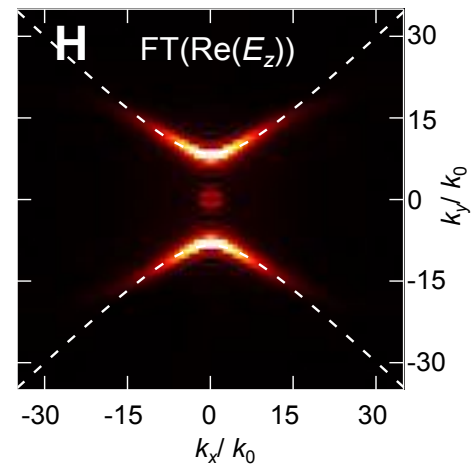
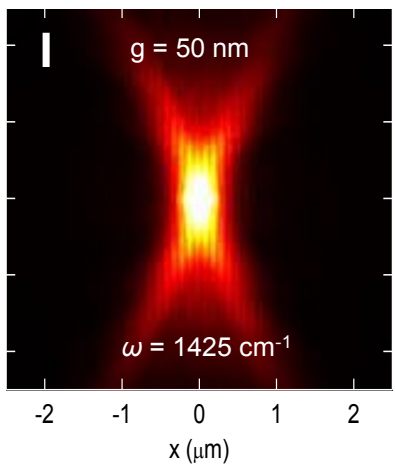 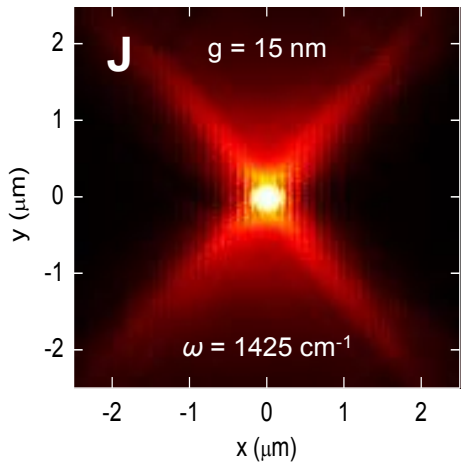 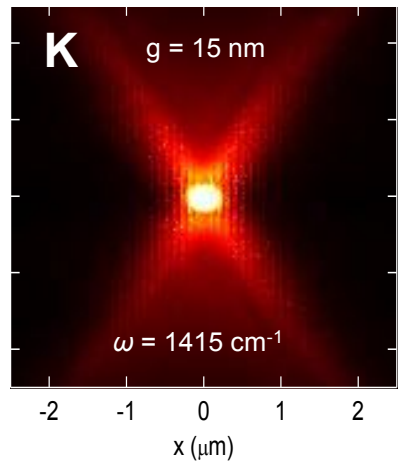 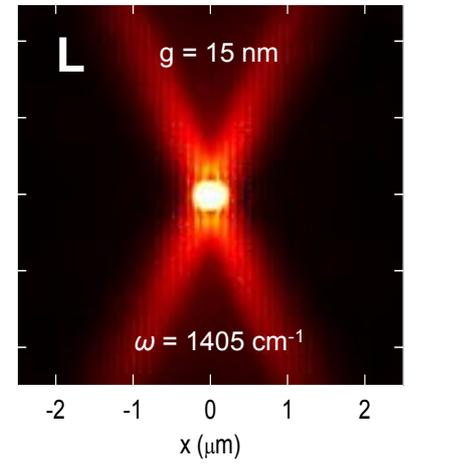

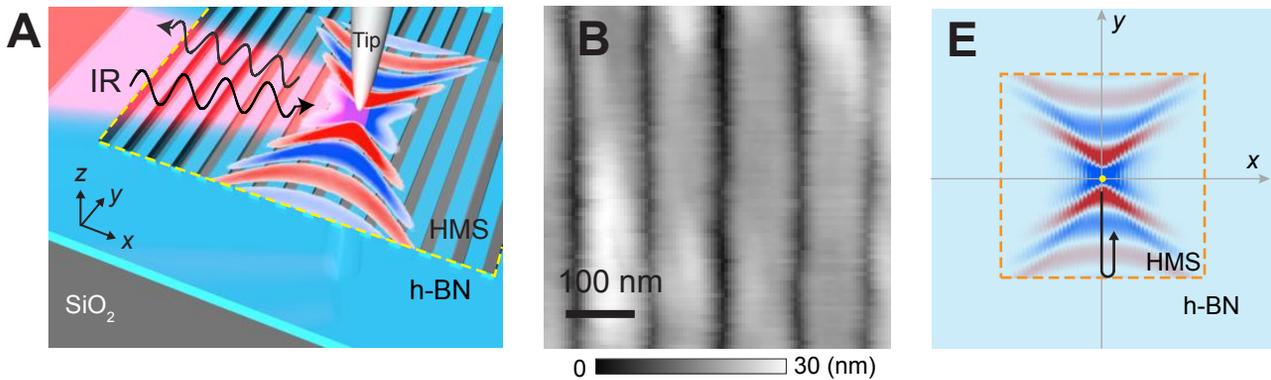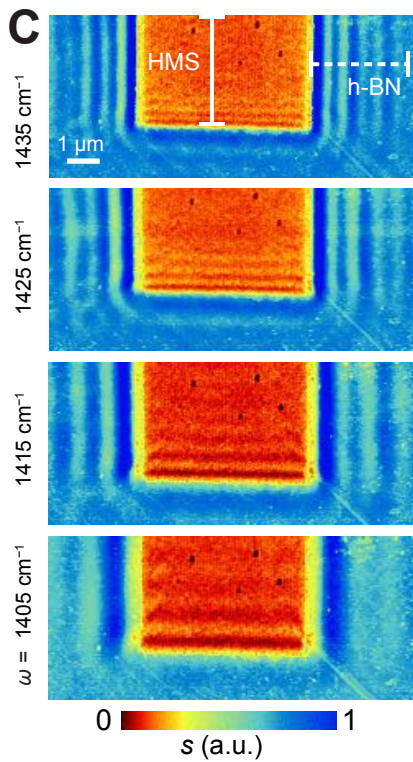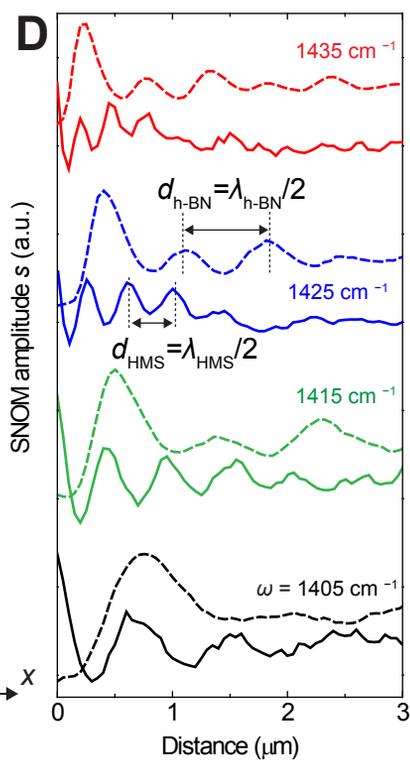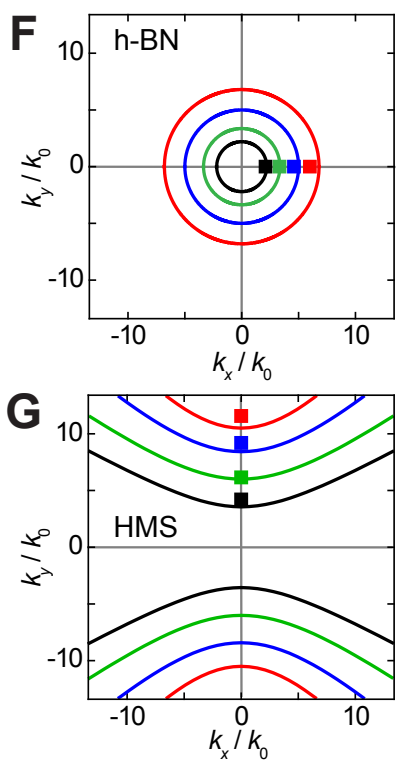

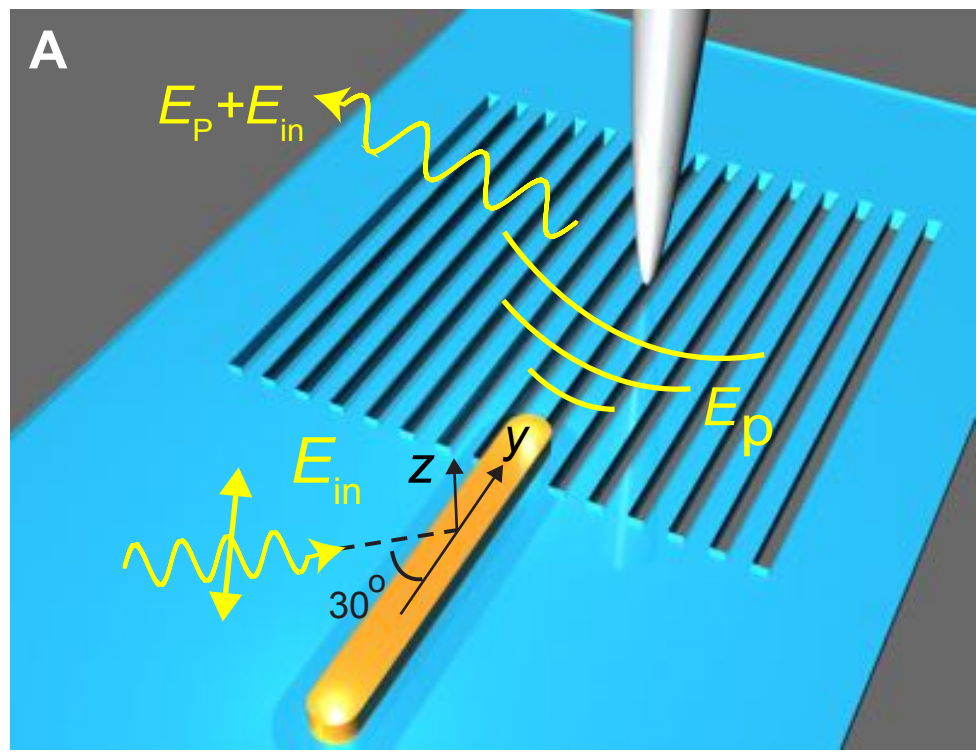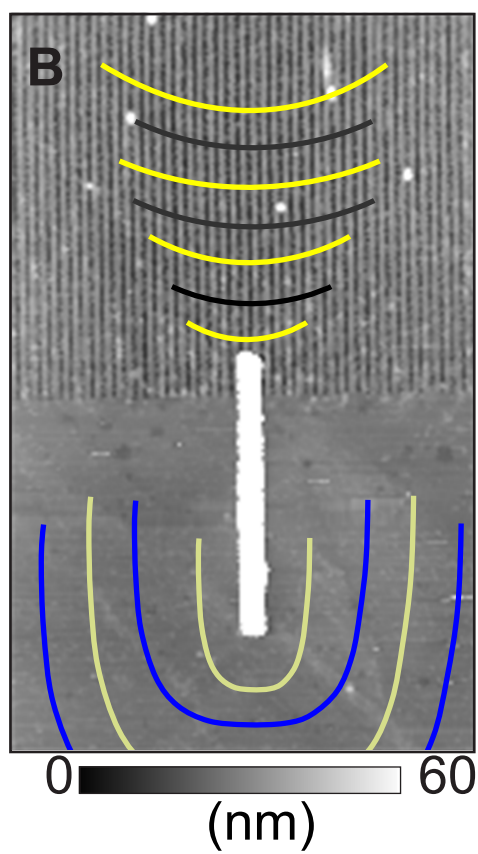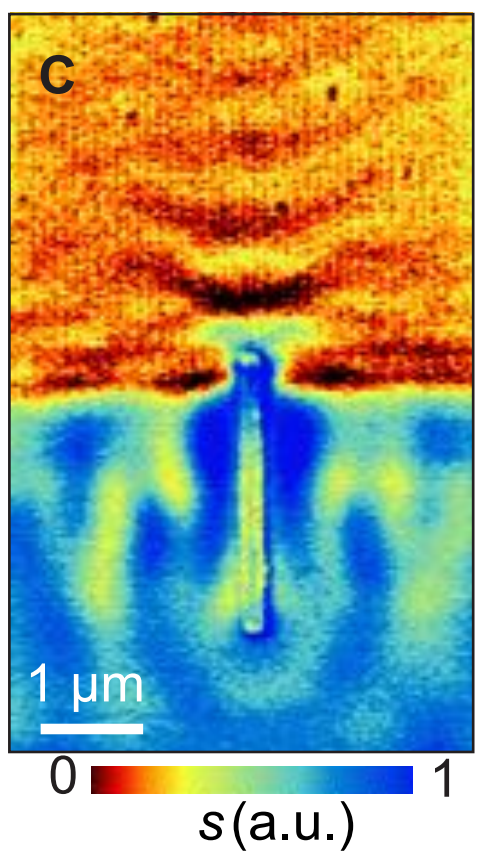

| A | B | C |
|---|---|---|
| $s$ (a.u.) 0 — 1 | $|E_z|$ (a.u.) 0 — 1 | $k_x / k_0$ −30 0 30 |

$\omega = 1435$ cm$^{-1}$

Experiment

Simulation

1 µm

1425 cm$^{-1}$

1415 cm$^{-1}$

$k_y / k_0$: 30, 15, 0, −15, −30